\documentstyle[epsfig]{ioplppt}

\def \K {$K^+/K^-$}
\def \fB {$\langle f \rangle_{B}$}
\def \np {$N_{\rm p}$}
\def \lp {$\bar{\Lambda}/\bar{p}$}
\def \Lp {$(\bar{\Lambda}+\overline{\Sigma^0}+1.1\overline{\Sigma^+})/\bar{p}$}
\def \lbar {$\bar{\Lambda}$}
\def \pbar {$\bar{p}$}

\begin{document}
\title{Strangeness in dense nuclear matter: A review of AGS results}
\author{Fuqiang Wang}
\address{Department of Physics, Purdue University, 1396 Physics, 
	West Lafayette, IN 47907\\
	Nuclear Science Division, Lawrence Berkeley National Lab, 
	Berkeley, CA 94720}

\begin{abstract}
Enhancement in strangeness production and antihyperon to antibaryon
ratio are two important signatures of QGP formation. Both signatures 
have been measured at the AGS. A review of these measurements is given.
Implications of the measurements are discussed.
\end{abstract}


\section{Introduction}

Nuclear matter at high baryon and/or energy density has been 
extensively studied through high energy heavy-ion collisions~\cite{qm}.
The primary goal of these studies is to observe the possible phase 
transition from hadronic matter to quark-gluon plasma (QGP),
in which quarks and gluons are deconfined over an extended 
region~\cite{Lee1,Lee2}. Enhancement in strangeness production
and antihyperon to antibaryon ratio are two of the proposed 
signatures of QGP formation. The underlying physics are as follows.

\begin{enumerate}
\item In a deconfined QGP, strange quark pairs ($s\bar{s}$)
can be copiously produced through gluon-gluon fusion
($gg \rightarrow s\bar{s}$)~\cite
{Rafelski1,Rafelski2,Rafelski3,Egg91:QGP,Kap86:cs},
while in a hadronic gas, $s\bar{s}$ pairs have to be produced via 
pairs of strange hadrons with higher production thresholds.
Moreover, the time scale of the gluon-gluon fusion process is short, 
on the order of 1--3~fm~\cite{Egg91:QGP}.
On the other hand, it is argued that strangeness production rate in a 
chemically equilibrated hadronic gas might be as high as that in a 
QGP~\cite{Kap86:cs,Lee88:prc}.
However, the time needed for a hadronic gas system to reach chemical 
equilibrium is significantly longer~\cite{Egg91:QGP,Kap86:cs,Lee88:prc}
than the typical life time of a heavy-ion reaction of the order of 10~fm.

\item In addition to enhanced production of strangeness, 
production of antihyperons should be further enhanced in 
QGP under finite baryon density~\cite{Lee88:prc,Ko92:prc}. 
This can be seen in the simple Fermi energy-level picture:
Low energy levels of light quarks ($q$) are already occupied
by the excessive light quarks due to the finite baryon density;
when the Fermi energies of light quarks are higher than the bare mass 
of a $s\overline{s}$ pair, $s\overline{s}$ pair production is 
energetically more favorable than that of $q\overline{q}$. 
Hence, the production of nonstrange light antiquarks is suppressed,
resulting in a high $\overline{s}/\overline{q}$ ratio in QGP.
\end{enumerate}

The definition of strangeness enhancement is broad. 
In this article, we use the following definition:
enhancement of strangeness production rate and antilambda to 
antiproton ratio in central heavy-ion collisions with respect 
to peripheral collisions, and to isospin weighted 
nucleon-nucleon (NN) interactions at the same energy.

The article is organized as follows.
In section~\ref{kaon}, we discuss experimental results on
the enhancement of kaon production at the AGS. 
In section~\ref{ratio}, we examine experimental results on
the average baryon phase-space density reached at the AGS, 
and discuss connections between kaon production and 
the average baryon phase-space density.
In section~\ref{antibaryon}, we discuss experimental results on
the enhancement of antilambda to antiproton ratio, 
and investigate the effect of high baryon phase-space density 
on the ratio. 
Finally, we draw conclusions in section~\ref{conclusion}.

\section{Kaon production enhancement\label{kaon}}

At AGS energies, the dominant carriers of strangeness produced in 
heavy-ion collisions are kaons (charged and neutral) and 
hyperons ($\Lambda$ and $\Sigma$'s).
Total charged and total neutral kaon yields are approximately equal 
under charge (isospin) symmetry. 
Hyperons are mostly produced through associated production with kaons.
Therefore, charged kaon yields give a fairly good overall scale of
strangeness production at the AGS.

\begin{figure}[hbt]
\hfill
\epsfig{file=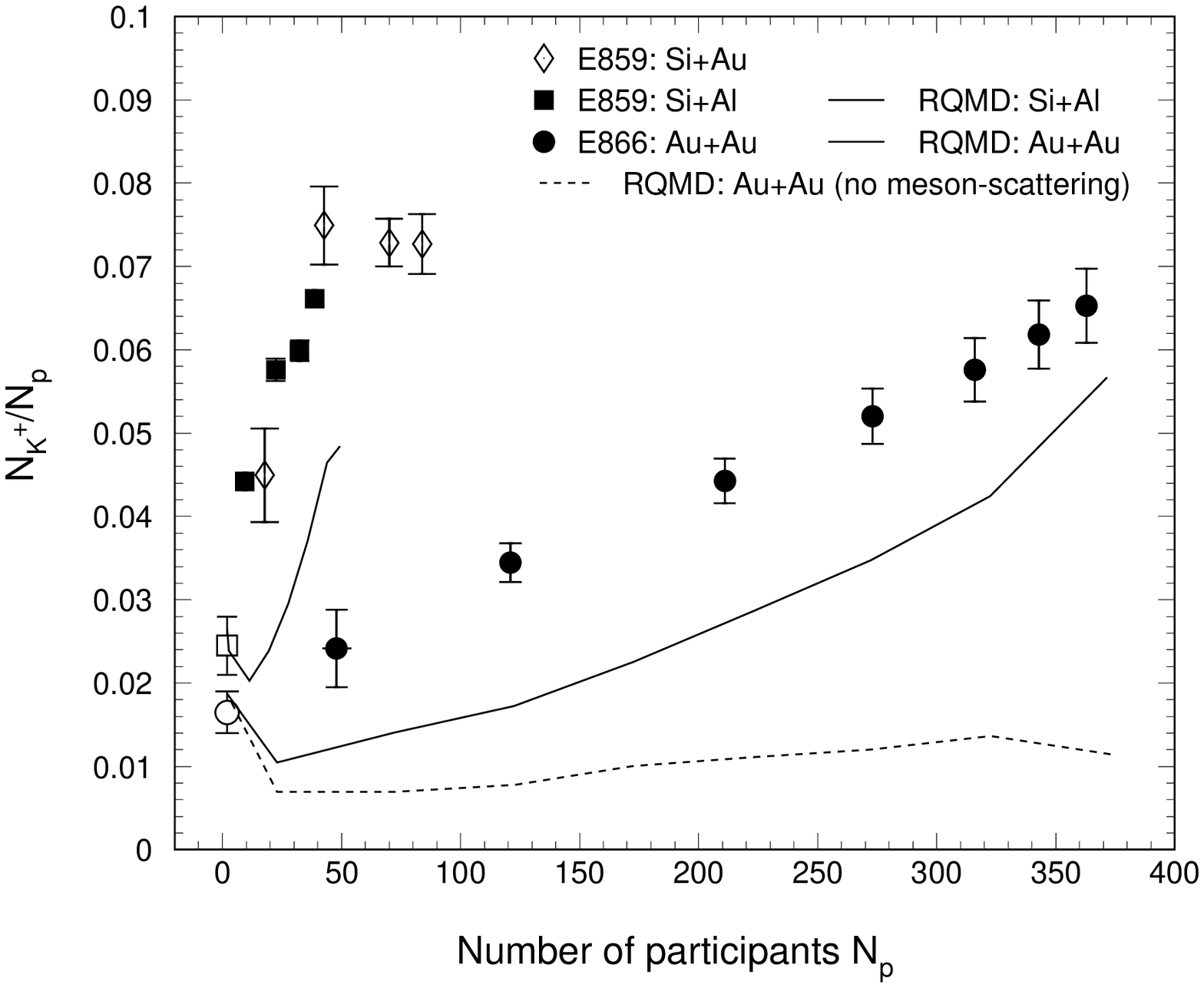,width=0.45\textwidth}
\epsfig{file=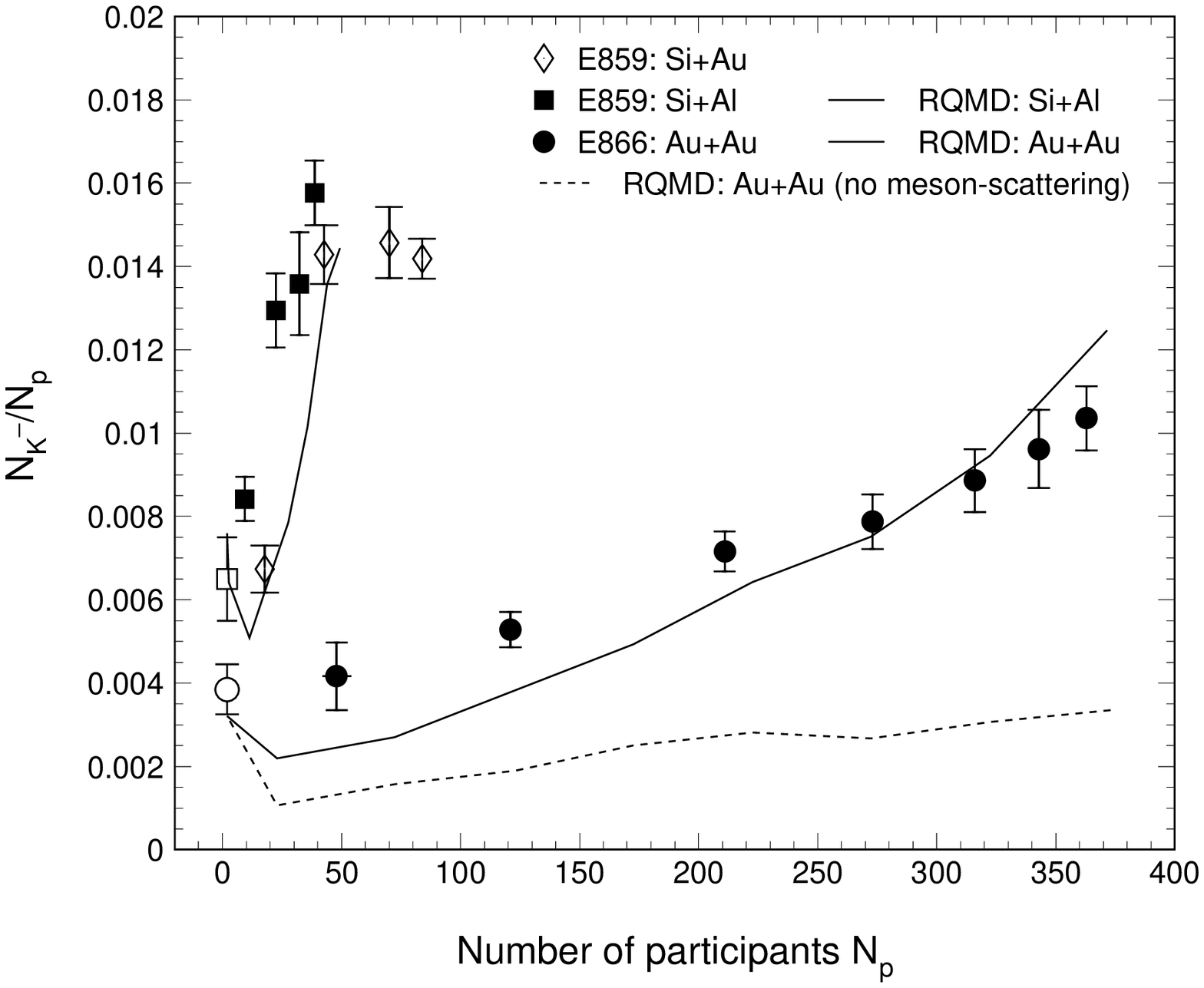,width=0.45\textwidth}
\caption{$K^+$ (left panel) and $K^-$ (right panel) yields per
participant as a function of the number of participants (\np)
for Si+A ($\sqrt{s}=5.39$~GeV) and Au+Au ($\sqrt{s}=4.74$~GeV) 
collisions at the AGS.
Those for isospin weighted NN interactions at $\sqrt{s}=5.39$~GeV
(open square) and $4.74$~GeV (open circles) are also shown.
RQMD results are shown in solid (the default settings) and dashed
curves (meson scattering switched off).
The left end of each curve is the corresponding RQMD result
for isospin weighted NN interaction.
The kaon yields for both data and RQMD
include $\phi$ feeddowns, which according to RQMD
can be neglected for $K^+$ and increase the $K^-$ yield by 10--15\%
(the default settings).}
\label{fig1}
\end{figure}

Kaon yields are systematically measured by AGS E802/859/866.
Figure~\ref{fig1} shows, in data points, kaon yields per participant
as a function of the number of participants (\np) in Si+A and 
Au+Au collisions~\cite{Ahl99:kaons}.
The data show that kaon production rate steadily increases with collision
centrality in Si+Al and Au+Au collisions.
For comparison, kaon yields per participant for isospin weighted
NN interactions at the corresponding energies are shown in
Fig.\ref{fig1} as the open circles and squares.
The enhancement factor -- the ratio of kaon yields per participant
in heavy-ion collisions over the same energy NN interactions -- 
is higher in Au+Au central collisions than in Si+A for both 
$K^+$ and $K^-$~\cite{Ahl99:kaons}.

The difference between Si+A and Au+Au at the same number of participants
(say \np\ $\sim 50$) is qualitatively consistent with the different
collision geometries under a binary collision model for kaon production. 
However, the difference cannot be quantitatively explained by 
Glauber-type calculations
of the average number of binary NN collisions~\cite{Ahl99:kaons}.

From a classical Glauber-type picture, in which each nucleon collides
multiple times with others in a heavy-ion collision, 
with each collision having certain probability producing kaons,
one would expect higher kaon production rate in heavy-ion collisions 
than NN.
To check this expectation, Figure~\ref{fig1} shows, 
in the dashed curves, results of the Relativistic Quantum Molecular
Dynamics (RQMD) model calculations for Au+Au 
collisions with meson-baryon and meson-meson interactions switched off.
One observes only a slight increase in kaon production rate from 
peripheral to central collisions.
In contrast to the expectation, the calculated kaon production
rate in heavy-ion collisions is lower than the corresponding NN value. 
This is probably due to the net effect of the following:
(1) not all the initial nucleons collide at least once at full energy;
(2) subsequent NN collisions have less than the full energy;
(3) AGS energy is near kaon production threshold where kaon production 
rate changes rapidly with energy in NN interactions~\cite{Ros75:pp}.

By including secondary meson-baryon and meson-meson interactions 
(the default settings), RQMD is able to describe the qualitative
features seen in the data. The results of the default calculations
are shown in Fig.\ref{fig1} as solid curves.
Within the RQMD model, both baryon-baryon and meson-induced
interactions increase kaon production rate. 
However, as seen from comparison between the dashed and 
solid curve (for Au+Au) in Fig.\ref{fig1}, 
the meson-induced interactions dominate for kaon production,
especially in central collisions~\cite{Sor91:meson}.
Although the default RQMD results seem to capture the qualitative 
feature of the data, quantitative differences exist between the data 
and the model. 
These differences warrant further studies~\cite{Soltz}.

\section{The average baryon phase-space density\label{ratio}}

It is obvious from Fig.~\ref{fig1} that, although the kaon production
rate increases with collision centrality reaching a factor of 3--4 
enhancement in central collisions with respect to NN interactions,
the ratio of charged kaon total yields (\K) 
varies little with the collision centrality~\cite{Ahl99:kaons}.
Similar results have been also observed in Ni+Ni collisions at the 
SIS~\cite{KaoS_NiNi} and Pb+Pb collisions at the SPS~\cite{Sikler}.
The nearly constant \K\ is puzzling because $K^+$ and $K^-$ are 
thought to be produced by different mechanisms: $K^-$'s are produced 
by pair production together with a $K^+$, while $K^+$'s can be produced, 
in addition, by associated production together with a hyperon. 
These different production mechanisms lead to different rapidity 
distributions which are observed, namely, the $K^+$ rapidity 
distribution is broader than $K^-$'s~\cite{Ahl99:kaons}. 
One naively expects that in heavy-ion collisions the relative 
contribution of associated over pair production increases with centrality, 
because the associated production threshold for kaons is lower than 
the pair production threshold, and there are more particle re-interactions 
(at lower than the full energy) in central than peripheral collisions. 
Therefore, one expects an increasing \K\ with centrality.

On the other hand, the constituent quark model~\cite{Anisovich,Bjorken} 
has been successfully applied to describe particle ratios in heavy-ion 
collisions~\cite{Bialas,Zimanyi}.
In the constituent quark model, 
$K^+ = u\overline{s}$ and $K^- = \overline{u}s$. 
Hence, \K\ depends on the baryon (baryon$-$antibaryon) phase-space 
density established in the collision zone at chemical freeze-out. 
In this picture, the observed constant \K\ implies a constant baryon 
phase-space density over the collision centrality.

\begin{figure}[hbt]
\hfill\epsfig{file=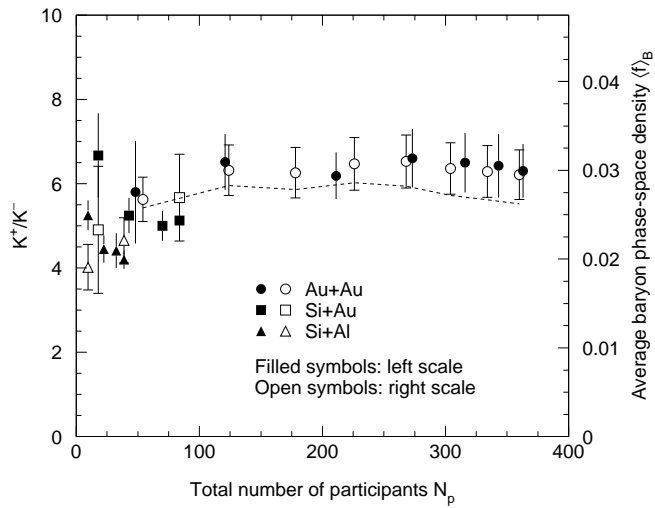,width=0.7\textwidth}
\caption{The \K\ ratio (filled symbols, left scale) and the average 
	baryon phase-space density at kinetic freeze-out \fB\ 
	(open symbols, right scale) as a function of the 
	total number of participants (\np) at the AGS.
	Errors shown are statistical for \K\ and statistical and
	systematic errors added in quadrature for \fB.
	The experimental systematic effects largely
	cancel in the \K\ and d/p ratios.
	Note that both quantities are nearly constant over centrality.
	The dashed curve shows the average nucleon phase-space density 
	at kinetic freeze-out (right scale); 
	the errors are comparable to the \fB\ ones.}
\label{fig2}
\end{figure}

The correlation between the kaon ratio and the average baryon phase-space
density at kinetic freeze-out (\fB) is examined in Ref.~\cite{kaonRatio}. 
The \fB\ values are experimentally determined from the deuteron to proton 
ratio (d/p) in the deuteron coalescence model.
Figure~\ref{fig2} shows \K\ and \fB\ at the AGS as a function of \np. 
Both quantities are nearly constant, and are consistent within the
constituent quark model.
Because \K\ is fixed at chemical freeze-out, the results may imply that 
the \fB\ values at kinetic freeze-out, which presumably happens later 
than chemical freeze-out in heavy-ion collisions, are connected to the 
chemical freeze-out values in a way that is independent of centrality.
Since baryons at AGS energies mostly consist of nucleons, 
a similar picture can be also made for the average nucleon phase-space
density, which is shown in the dashed curve in Fig.~\ref{fig2}.

\begin{figure}[hbt]
\hfill
\epsfig{file=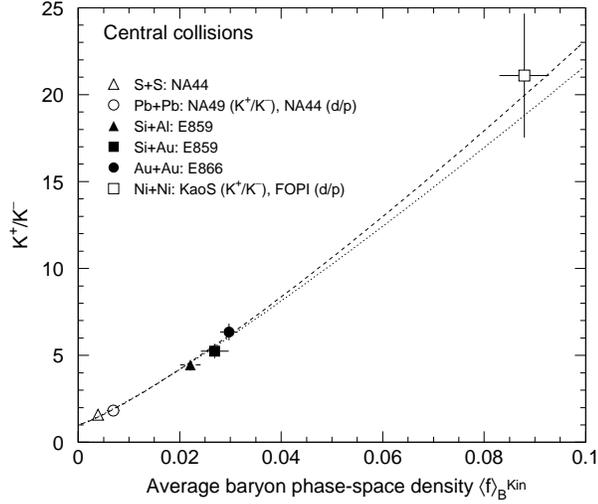,width=0.65\textwidth}
\caption{The \K\ ratio as a function of the average baryon phase-space density 
	at kinetic freeze-out \fB\ in central heavy-ion collisions.
	Errors shown are statistical and systematic errors added in 
	quadrature. The dashed curve is a fit to power-law function. 
	The dotted curve is a similar fit but excluding the Ni+Ni data point.}
\label{fig3}
\end{figure}

The correlation between \K\ and \fB\ in central collisions at 
different beam energies has been also studied in Ref.~\cite{kaonRatio}.
Figure~\ref{fig3} shows \K\ as a function of \fB.
Both quantities decrease with increasing beam energy. 
The data from various collision systems at different beam energies 
follow a similar dependence. This dependence can be parameterized 
by a power-law function as motivated by equilibrium models. 
The result of such a parameterization is shown in the dashed curve.

These results may have implications on medium modification to the effective
kaon mass. In fact, the KaoS Collaboration has inferred a significant
$K^-$ mass drop from their \K\ measurement.
Since \fB\ is a measure of matter density, 
studying \K\ against \fB\ may provide a more direct 
approach to search for medium effect on kaon mass: 
If the $K^-$ mass is modified by medium at low energy
but not (or differently) at high energy, 
then the low energy \K\ should 
deviate from extrapolation of the high energy data.
The common systematic in Fig.~\ref{fig3} seems to imply no difference 
between the SIS and AGS/SPS data.
However, one should be careful to conclude from these results
that there is {\em no} $K^-$ mass drop in the Ni+Ni data. 
This is because: 
\begin{enumerate}
\item 
Although the power law parameterization is motivated
by equilibrium models, it is not a priori for the relation
between \K\ and \fB.
\item
Since the leverage of the AGS and SPS data is small, the systematic
errors on these data are critically important to the extrapolation to the
SIS data. 
\end{enumerate}
In order to address these concerns, more data are needed to fill
in the gap between AGS and SIS.
These data are underway from the E895 and the E917 Collaborations.
New data from the SPS at 40 AGeV would be also valuable 
as they fill in the gap between the full-energy AGS and SPS data.

\section{The antilambda to antiproton ratio\label{antibaryon}}

The AGS/E864 Collaboration has deduced a \Lp\ ratio 
in Au+Au collisions at mid-rapidity and zero $p_T$ by attributing
the discrepancy between their \pbar\ measurement and that from AGS/E878 
entirely to the different acceptances of the two experiments
for \pbar's from weak decays~\cite{E864PRC}.
The deduced \Lp\ value, while being consistent with p+p results 
in peripheral collisions at similar energies, 
has a strong dependence on the collision centrality.
In the most 10\% central collisions, the \Lp\ ratio reaches a most
probable value of 3.5, which is over an order of magnitude 
larger than in p+p interactions.
The E917 Collaboration has made direct measurements of \pbar\ and
\lbar\ yields at mid-rapidity and integrated over $p_T$~\cite{E917}.
The ratio of \lp, given the large error bar, is consistent with E864.
These results are intriguing because they may point to possible
QGP formation as discussed in the introduction.

There are at least two physical origins for the large \Lp\ ratio: 
(1) an enhanced \lp\ (and/or $\bar{\Sigma^{0,+}}/\bar{p}$) ratio
at the initial production stage of the antibaryons; and 
(2) a strong absorption of \pbar's and a less strong absorption 
of \lbar's and $\bar{\Sigma}$'s
in nuclear matter produced in heavy ion collisions.
An enhanced \lp\ ratio at the initial production stage would be 
an evidence for QGP formation. In order to obtain the \lp\ ratio
at the initial stage, one has to postulate from measurements at
the final freeze-out stage including nuclear absorption effects.
To this end, we use the Ultra-relativistic Quantum Molecular Dynamics 
(UrQMD) model~\cite{Bass} to simulate Au+Au collisions at the AGS 
and record the initial production abundances of antibaryons and high
mass antibaryon resonances as well as the final freeze-out abundances. 
We chose UrQMD because 
(1) it has been reasonably successful in describing many of the 
experimental results on hadron spectra, as well as 
the average baryon density and the baryon emitting source size
which are the essential ingredients for nuclear absorption, and 
(2) it does not have a QGP state or mechanisms mimicking a QGP state.
Our strategy is then to compare the final freeze-out \Lp\ ratio to
data and use the initial \lp\ information from UrQMD to conjecture 
what the data might be telling us.

Once produced, an antibaryon may or may not annihilate with baryons
in the collision zone.
The $\bar{p} p$ annihilation cross section is well measured, and is used
in UrQMD. In this study, the $\overline p p$ annihilation 
cross section is given by
\begin{equation}
\sigma_{\bar{p}p}^{\rm ann}(\sqrt s) 
= 1.2 \; {\rm GeV} \cdot
\frac{\sigma_{\bar{p}p}^{\rm tot}(\sqrt s)}{\sqrt s}\quad .
\end{equation}
The total $\overline p p$ cross section, $\sigma_{\bar{p}p}^{\rm tot}$,
is taken from the CERN-HERA parameterization.
The other antibaryon-baryon annihilation cross sections are,
however, not well measured.
UrQMD applies a correction factor, given by the Additive Quark Model 
for these annihilation cross sections~\cite{bleicherReview}:
\begin{equation}
\frac{\sigma_{\bar{B}B}(\sqrt s)}{\sigma_{\bar{p}p}(\sqrt s)} = 
\left(1-0.4\frac{s_{\bar{B}}}{3}\right)
\left(1-0.4\frac{s_B}{3}\right).
\label{eq:aqm}
\end{equation}
Here $s_B$ and $s_{\bar{B}}$ are the strangeness number of the 
baryon and the antibaryon, respectively.
There is a reduction of 40\% due to each strange or anti-strange quark.
For instance, the $\bar{\Lambda}p$ annihilation cross section is
\begin{equation}
\sigma_{\bar{\Lambda}p}(\sqrt s) = 0.87 \sigma_{\bar{p}p}(\sqrt s).
\label{xsection}
\end{equation}
Note that the above relation is for the same center-of-mass (c.m.s)
energy, $\sqrt s$, of $\bar{\Lambda}p$ and $\bar{p}p$ systems. 
Generally, the $\bar{\Lambda}p$ c.m.s. energy is
larger than that of $\bar{p}p$ in heavy ion collisions, 
so the average reduction factor should be lower than 0.87 
as $\sigma_{\bar{p}p}(\sqrt s)$ is a decreasing function of $\sqrt{s}$.

\begin{figure}
\hfill\epsfig{file=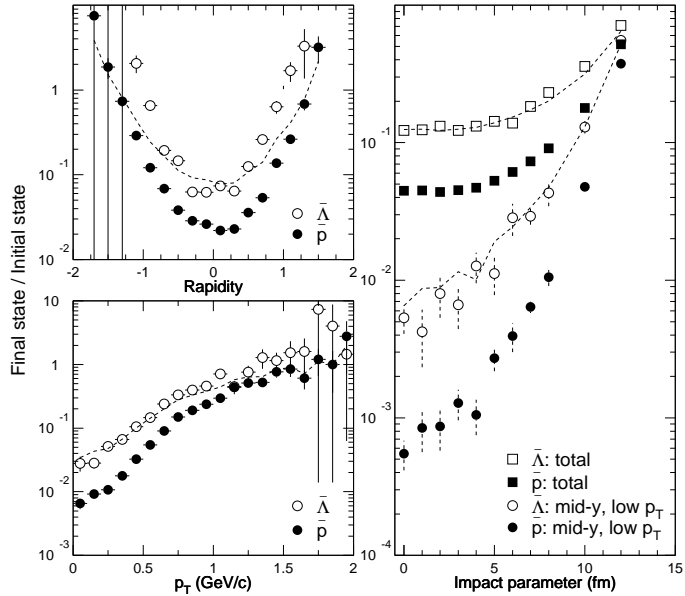,width=0.7\textwidth}
\caption{The ratio of number of \pbar's (solid symbols) 
and \lbar's (open symbols) at final freeze-out over that at initial 
string fragmentation stage, calculated by UrQMD for central Au+Au collisions
($b<1$ fm), as a function of 
(a) rapidity but integrated over $p_T$, and
(b) $p_T$ but integrated over rapidity, 
and for all collisions, as a function of
(c) impact parameter for mid-rapidity ($|\Delta y|<0.4$) and low $p_T$ 
($p_T<300$ MeV/c) region (circles)
as well as for integrated over whole phase-space (squares).
At AGS energies, antibaryons can be produced only at the very early time 
from string fragmentation, and are then annihilated by baryons at later
times. The ratio can be larger than one because the final 
freeze-out rapidity and/or $p_T$ distributions can be broader than
the initial distributions.
With this caution, the ratio may be viewed as the ``survival probability''
of \pbar\ and \lbar\ through the nuclear matter created in the collisions.
See text for explanations of the dashed curves.}
\label{ann}
\end{figure}

Figure~\ref{ann} gives an idea about the magnitude of the absorption
effect in Au+Au collisions by plotting the ratio of freeze-out \pbar\
(or \lbar) over that at the initial production stage.
The ratio can be viewed as the ``survival probability'' 
of \pbar\ (or \lbar) from initially produced to the final freeze-out.
The ratio is higher than one in forward and backward rapidities
because the final rapidity distributions can be broader than the initial ones.
The left upper plot shows the ``survival probability'' as a function of 
rapidity, the left lower plot as a function of $p_T$ for central 
Au+Au collisions.
It is clearly seen that the largest absorption occurs in the mid-rapidity
and low $p_T$ region.
The right panel shows the ``survival probabilities'' of mid-rapidity and 
low $p_T$ \pbar's and \lbar's in solid symbols and integrated over
whole phase-space in open symbols, as a function of impact parameter in 
Au+Au collisions. 
As seen from the plot, about 99.9\% and 99\% of the mid-rapidity and 
low $p_T$ \pbar's and \lbar's produced in central Au+Au collisions
are annihilated according to UrQMD. In other words, only 1
out of 1000 \pbar's and 1 out of 100 \lbar's in this kinematic region
survive to freeze-out.

On the other hand, if all the \pbar's and \lbar's are counted, 
then the ``survival probabilities'' are much higher, and are roughly
constant over a wide range of impact parameter in central collisions.
However, this has implications on the interpretations of the
measured absolute \pbar\ yields.
The measured rapidity density (within a fixed rapidity window) 
has a less than linear increase with \np\ in Au+Au 
collisions~\cite{E866pbar}. The power factor of the increase is 0.74.
If we take into account the absorption effect shown in the filled squares
in Fig.~\ref{ann}(c), then the restored rapidity density of initially 
produced \pbar's would have a stronger than linear increase with \np\
participants, to a value of about 0.4 in central collisions, 
and the power factor would be about 1.5.

Now back to Fig.~\ref{ann}.
In the simple picture of a sphere of baryons with a uniform density $\rho$
and radius $R$, the survival probability of an antibaryon produced at center
is $\exp(-\sigma^{\rm ann}\rho R)$, where $\sigma^{\rm ann}$ is 
the annihilation cross section.
In order to obtain an intuitive idea, one may recall Eq.~\ref{xsection} 
and scale the \pbar\ results by a power factor smaller than 0.87 to
obtain the \lbar\ survival probability. 
This is done in the dashed curves in Fig.~\ref{ann}: 
They are the \pbar\ points to the power 2/3 (an {\it ad hoc} number). 
It is worthwhile to note at this point that, although the average
baryon phase-space density (hence $\rho$) is fairly constant over 
centrality, the size $R$ strongly increases with centrality, 
resulting in a strong decrease in the survival probability of antibaryons.

\begin{figure}
\hfill
\epsfig{file=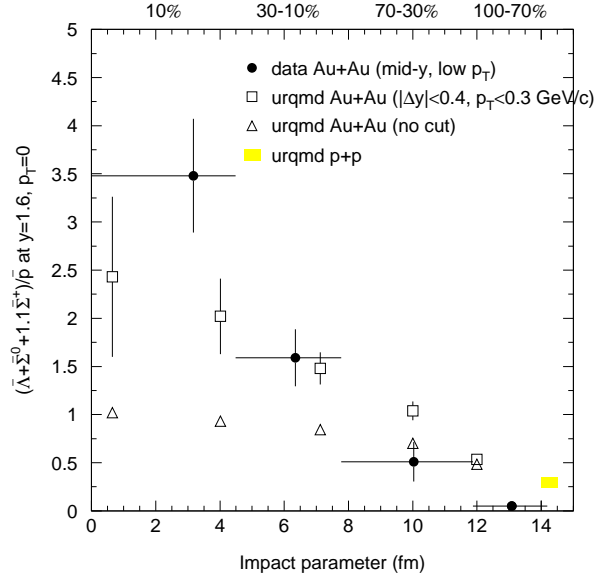,width=0.7\textwidth}
\caption{UrQMD calculation of the freeze-out \Lp\ ratio at mid-rapidity
($|\Delta y|<0.4)$ and low $p_T$ ($p_T<300$ MeV/c) as a function of impact
parameter (open squares), compared to experimental results in similar
kinematic region (filled circles). 
The experimental centrality bins are indicated above the top axis.
The UrQMD total yield \Lp\ ratio (no kinematic cut)
is shown in open triangles. The UrQMD calculated \Lp\ ratio 
in p+p interactions is shown in the shaded area, artificially plotted
at a large impact parameter value.
Note that UrQMD can reasonably reproduce the data.}
\label{freezeout}
\end{figure}

Figure~\ref{freezeout} shows the freeze-out ratio of \Lp\ at mid-rapidity
($|y|<0.4$ and $p_T<0.3$~GeV/c) in Au+Au collisions at AGS energy as a 
function of the collision impact parameter. The experimentally deduced ratio
is reproduced from Ref.~\cite{E864PRC} with impact parameter values obtained
from the centrality bins. 
The UrQMD ratio is in a good agreement with the data.
The \Lp\ ratio of the total particle yields are also shown. 
It is clear that the large ratio at mid-rapidity and low $p_T$ region 
in central collisions is largely due to the kinematic cut. 
Note that the calculated \Lp\ ratio in p+p interactions is consistent
with the trend calculated for heavy ion collisions.

As UrQMD successfully describe the data, it is interesting to examine
the \lp\ ratio at the initial production stage.
This is because a significantly larger value of this ratio in heavy 
ion collisions than in N+N interactions may signal QGP formation.
The extracted initial \lp\ ratio is independent on centrality, 
and is consistent with experimental measurement of about 0.2. 
This is reasonable because string fragmentation function
shall not know about the centrality of the collisions, and
there is no QGP in UrQMD. 
As UrQMD well reproduces the freeze-out values for the \Lp\ ratio, 
it is reasonable to suggest that the data indicate no large ratio of \lp\ 
at the initial stage, therefore do not exclusively imply a QGP formation.

\section{Conclusions\label{conclusion}}

We have reviewed the AGS results on kaon production, antilambda to antiproton
ratio, and the role of the large average baryon phase-space density.
We draw the following conclusions:
\begin{enumerate}
\item Enhancement in charged kaon production is observed in central 
	heavy-ion collisions at the AGS with respect to peripheral 
	collisions and isospin weighted nucleon-nucleon interactions. 
	The enhancement can be qualitatively described by secondary 
	particle scattering within the RQMD model.
	However, quantitative differences exist between the data and RQMD,
	which warrant further investigation.
\item The average baryon phase-space density is experimentally 
	extracted and is found to be constant over the collision 
	centrality at the AGS.
	This explains, within the constituent quark model, 
	the observed constant charged kaon ratio. 
	The average baryon phase-space density is further
	studied for central collisions and is found to increase 
	with decreasing beam energy. The charged kaon ratio is
	found to strongly correlate with the average baryon phase-space
	density over a wide range of beam energy.
\item Strong increase of the \lp\ ratio from peripheral to central 
	collisions is deduced from experimental data at the AGS.
	The hadronic transport model, UrQMD, can satisfactorily 
	describe the large ratio in central collisions and the 
	centrality dependence of the ratio. According to the model,
	the experimentally deduced large ratio of \lp\ in the mid-rapidity
	and low $p_T$ region is mainly due to the strong absorption
	of these \pbar's and \lbar's.
	Therefore, the data do not exclusively imply a QGP formation.
\end{enumerate}

\section*{Acknowledgments}

I am especially grateful to Dr. Grazyna Odyniec for 
inviting me to review the AGS experimental results.
It was also a delight to return to Berkeley after a short leave.
The work on the absorption effect on the antilambda to antiproton ratio
was collaborated with Dr. M. Bleicher.
Thanks also go to Dr. U. Heinz and Dr. K. Redlich for fruitful discussions.
This work was supported by the U.S. Department of Energy
under contracts DE-FG02-88ER40412 and DE-AC03-76SF00098.

\end{document}